\documentclass[usegraphicx,useAMS]{mn2e}

\newcommand{\kms}{\mbox{${\rm km\,s}^{-1}$}}

\title[Kinematic structure in the $\sigma$ Ori association]
  {Kinematic structure in the young $\sigma$ Orionis association}
\author[R.D. Jeffries et al.]
  {R.D.~Jeffries$^{1}$, P.F.L.~Maxted$^{1}$, J.M.~Oliveira$^{1}$ and Tim Naylor$^{2}$\\
$^{1}$  Astrophysics Group, Keele University, Keele, 
      Staffordshire ST5 5BG, United Kingdom\\
$^2$ School of Physics, University of Exeter, Stocker Road, Exeter
  EX4 4QL, United Kingdom\\
}
\date{Submitted May 18th 2006}

\pagerange{\pageref{firstpage}--\pageref{lastpage}} \pubyear{2004}

\def\LaTeX{L\kern-.36em\raise.3ex\hbox{a}\kern-.15em
    T\kern-.1667em\lower.7ex\hbox{E}\kern-.125emX}

\begin{document}

\label{firstpage}

\maketitle

\begin{abstract}
We have used precise radial velocity measurements for a large number of
candidate low-mass stars and brown dwarfs, to show that the young
$\sigma$~Ori ``cluster'' consists of two spatially superimposed
components which are kinematically separated by 7\,\kms\ in
radial velocity, and which have different mean ages. We examine the
relationship of these two kinematic groups to other populations in the
Orion OB1 association and briefly discuss the consequence of mixed age
samples for ongoing investigations of the formation and evolution of
low-mass objects in this much-observed region.

\end{abstract}

\begin{keywords}
stars: low-mass, brown dwarfs -- stars: pre-main-sequence -- open clusters
and associations: individual: $\sigma$~Orionis.
\end{keywords}

\section{Introduction}

Since their discovery (Walter et al. 1997), the low-mass objects
surrounding the young star $\sigma$ Ori (a quintuple system with an
O9.5V primary and four spatially resolved, early-type components) have
become one of the most intensively studied populations of young stars
and brown dwarfs. They are reasonably close, so the young,
intrinsically luminous, low-mass objects are bright.  Interpretation of
results is further aided by a high galactic latitude and low extinction
(Brown, de Geus \& de Zeeuw 1994).

Examples of recent investigations include the search for very low-mass
stars, brown dwarfs and even ``free floating planets'' and
determination of the initial mass function at low masses (e.g.
Zapatero Osorio et al. 2000, 2002a; B\'ejar et al. 2001; Sherry, Walter
\& Wolk 2004); using the association to constrain the timescales for disc
dissipation and the cessation of accretion in low-mass objects
(e.g. Barrado y Navascu\'es et al. 2003; Oliveira et al. 2004);
searching for analogues of classical T-Tauri stars among young
substellar objects (e.g. Jayawardhana et al. 2003); studying the
evolution of magnetic activity and angular momentum in low-mass stars
and brown dwarfs (e.g. Franciosini, Pallavicini \& Sanz-Forcada 2006);
and, investigating variability and possible age spreads in young star
forming regions (e.g. Caballero et al. 2004; Burningham et al. 2005a).

Most of these investigations rely on the assumption of either a
distance or age for the low-mass objects in the
association. Furthermore, an implicit assumption often made, is that
the studied objects are coeval or at least have an age spread that is
modest compared with the cluster age. Previous papers have either
adopted the Hipparcos distance to $\sigma$ Ori ($352^{+166}_{-85}$\,pc
-- Perryman et al. 1997) and deduced an age of $\simeq 5$\,Myr from
low-mass model isochrones (e.g. Oliveira et al. 2004), or they have
assumed that $\sigma$ Ori is part of the larger Orion OB1b association
that has a mean Hipparcos distance of $\simeq 440$\,pc (Brown, Walter
\& Blauuw 1998) and hence deduced an age of $\simeq 2.5$\,Myr
(e.g. Sherry et al. 2004).

In this letter we present the initial results of a precise radial velocity
survey of low-mass stars and brown dwarfs around $\sigma$~Ori. We show
that the ``cluster'' of low mass object around $\sigma$~Ori actually
consists of two, kinematically distinct, spatially superimposed
populations, with different mean ages.

\section{Observations and Analysis}

\begin{figure}
\includegraphics[width=86mm]{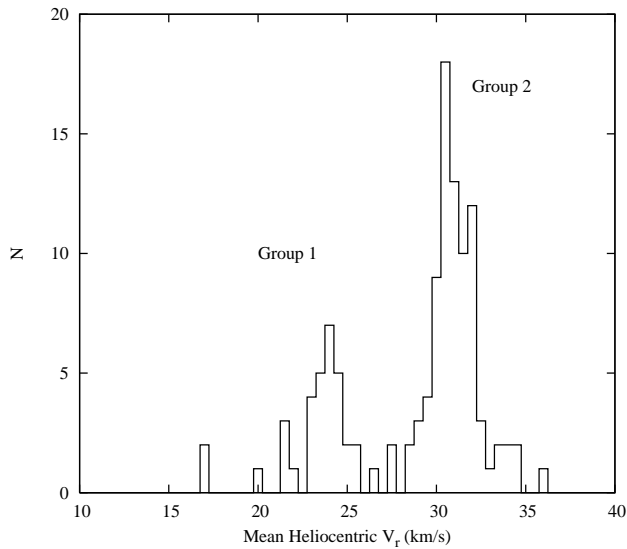}
 \caption{A histogram of heliocentric radial velocities for the
   filtered subsample described in the text. Two kinematic groups with
   small internal velocity dispersions are identified.
}
    \label{rvhist}
\end{figure}

\begin{figure}
\includegraphics[width=80mm,bb= 84 195 488 570, clip= ]{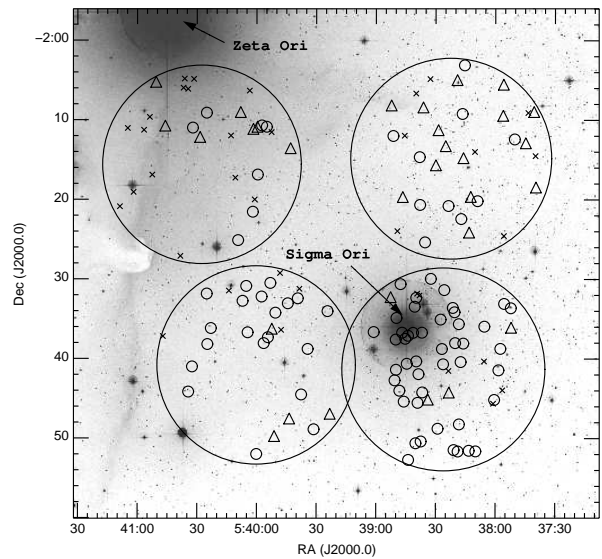}
 \caption{The spatial distribution of the filtered subsample described
 in the text. The two kinematic groups are indicated by triangles
 (group 1) or circles (group 2). Small crosses are targets 
 in neither kinematic group (wrong velocity, variable velocity or low
 EW[Na] -- see text) and the four fields observed with the Giraffe
 spectrograph are indicated with large circles.
}
    \label{spatial}
\end{figure}

Multi-object spectroscopy was performed at three epochs for four
separate fields (see Fig.~\ref{spatial}) around $\sigma$~Ori during the
period 10 November 2005 to 13 January 2006. A total of 148 targets were
observed at each epoch using the FLAMES instrument mounted on the
VLT-Kueyen (UT2) 8.2-m telescope. The Medusa
fibre system fed the Giraffe spectrograph and a 316 lines\,mm$^{-1}$
grating produced spectra with a resolving power of 16000, covering the
wavelength range 8080--8630\AA\ at 0.14\AA\ per pixel.  Targets with
$14.1<I<19.3$ were selected from the $I$ versus $R-I$ colour-magnitude
diagram (CMD) described by Kenyon et al. (2005), corresponding to
masses of $0.5>M/M_{\odot}>0.03$ (for an age of 3\,Myr and distance of
440\,pc -- Baraffe et al. 1998, see Fig.~\ref{cmd}).

A single exposure of 2750\,s was obtained at each epoch, yielding
spectra with signal-to-noise ratios (SNR) of 5--100 per pixel. One or
two bright F/G stars were simultaneously observed in each field using
the FLAMES fibre feed to the UVES spectrograph. These gave very high
SNR spectra at a resolving power of 47000, which covered the same
wavelength range and allowed us to model telluric absorption.

Spectra were reduced and extracted using the FLAMES UVES and Giraffe
pipeline software. Giraffe wavelength calibration was initially
provided by thorium-argon lamp exposures and then improved by reference
to telluric absorption lines in each extracted spectrum. Sky was
subtracted by forming a clipped mean sky spectrum from the many fibers
allocated to sky in each configuration. Telluric correction was
provided by modelling telluric features in the simultaneous UVES
spectra using a 6-layer atmospheric model (Nicholls 1988) and the HITRAN molecular
database (Rothman et al. 2005).  This synthetic telluric spectrum was
convolved with a gaussian profile 
and divided into the lower resolution Giraffe spectra. This
removed all visible traces of telluric contamination.

Heliocentric radial velocities (RVs) were found by cross-correlation
with the template star HD\,34055 (M6V) using the wavelength range
8061--8535\AA, but masking out regions around sky emission
lines. Internal uncertainties in the RVs are $<1$\,\kms\ for most
targets. The zero point was set by cross-correlating HD\,34055 with
similar UVES spectra of the M-dwarfs Gl\,402, Gl\,406 and Gl\,876,
which have precisely measured heliocentric RVs (see Bailer-Jones
2004). The heliocentric RV of HD\,34055 was $+44.2$\kms, with a range
when using the three standards that leads to an external error estimate
of $\pm 0.5$\,\kms.

Equivalent widths (EWs) for the gravity sensitive Na\,{\sc i} (8183,
8195\AA) features were found by integrating in a $\pm 120$\kms\ range
around the line centres (corrected for the RV). For the following
analysis steps we then excluded: objects with a significantly variable
RV over the three epochs (the topic of binarity in the sample will be
tackled in a subsequent paper); all objects with an EW[Na]\,$<1$\AA\
(indicative of a giant spectral class -- see below); and all objects
with a constant RV outside the range $10<V_{r}<40$\,\kms. A histogram
of the weighted mean RVs for the remaining 117 objects is shown in
Fig.~\ref{rvhist}. The RV distribution is clearly bimodal and we refer
to these two populations as kinematic groups 1 ($20<V_{r}<27$\,\kms)
and 2 ($27<V_{r}<35$\,\kms) respectively.

\begin{figure}
\includegraphics[width=86mm]{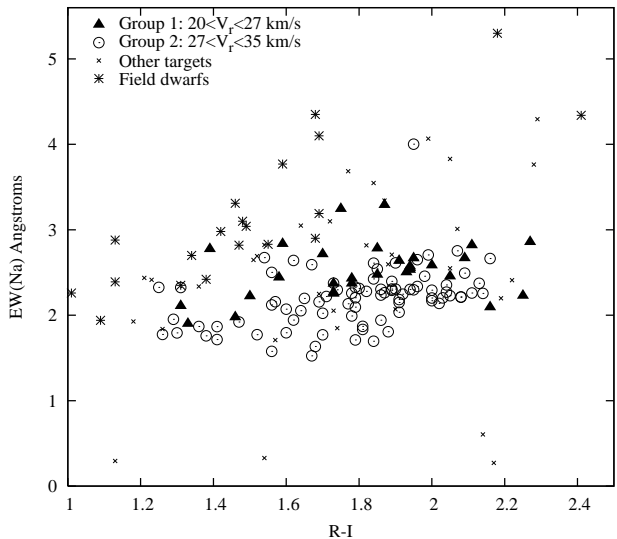}
 \caption{The total equivalent width of the Na\,{\sc i} (8183, 8195\AA)
 doublet as a function of $R-I$ colour. Members of the two kinematic
 groups are indicated (triangles -- group 1, circles -- group 2) and
 compared with data for field M dwarfs obtained with similar spectral
 resolution (Xu 1991; Montes \& Mart\'{i}n et al. 1998; Kenyon et
 al. 2005).  Small crosses mark targets in neither kinematic
 group.
}
    \label{na}
\end{figure}

\begin{figure}
\includegraphics[width=86mm]{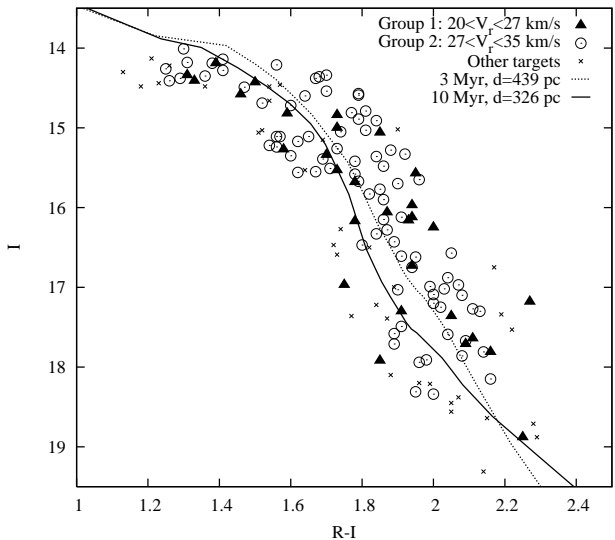}
 \caption{The colour magnitude diagram for the two kinematic
 groups (triangles -- group 1, circles -- group 2). Small crosses
 mark spectroscopic targets in neither kinematic group.
 Isochrones overplotted from Baraffe et al. (1998)
 are for  ages/distances of
 3\,Myr/439\,pc and 10\,Myr/326\,pc. Reddening and extinction
 corresponding to $E(R-I)=0.036$ has been applied.
}
    \label{cmd}
\end{figure}

Figure~\ref{spatial} shows the spatial distribution of objects in the
two kinematic groups.  Representatives of both groups are found in all
4 Giraffe fields (and at all epochs), but the ratio of group 2 to group
1 objects becomes smaller towards the north (e.g. this ratio is 10/15
in the north-west Giraffe field compared with 46/4 in the south-west
field).  Note that the photometric selection criteria for each of these
fields were identical. The larger total number of targets in the
south-west field simply reflects the larger number of photometric
candidates near $\sigma$~Ori -- a fact previously used to support the
existence of the ``$\sigma$~Ori cluster'' (e.g. Sherry et al. 2004).

Figure~\ref{na} shows how EW[Na] behaves with colour for each
group. This feature strengthens considerably with gravity in cool stars
(e.g. Schiavon et al. 1997). At young ages, low-mass stars and brown
dwarfs contract almost isothermally, resulting in a factor of $\simeq
10$ increase in gravity between 1 and 100\,Myr at a similar colour
(Baraffe et al. 1998).  Objects in kinematic groups 1 and 2 have a
weaker EW[Na] than field M dwarfs of similar colour but stronger than
giants where EW[Na] $\leq 1$\AA\ is found. Hence they have gravities
intermediate between dwarfs and giants as expected for young low-mass
objects. Figure~\ref{na} also shows that {\em on average} the objects
in group 2 have lower gravities than those in group 1, although there
is considerable overlap in the distributions. Formal EW uncertainties
are less than 0.1\AA\ for most of our targets and so cannot account for
the overlap. Qualitatively, this indicates that group 1 is older (on
average) than group 2, but the lack of any detailed EW-age calibration
prevent us from making a quantitative estimate of the age difference or
any age spread within each group.

Figure~\ref{cmd} shows the $I$ versus $R-I$ CMD for the two kinematic
groups as well as for all the other spectroscopic targets to illustrate
the target selection region.  No obvious difference in distribution is
seen. Of course all the targets were selected using the same
photometric criteria, but Burningham et al. (2005b) have shown that few
young stars exist outside of this range near $\sigma$~Ori. Therefore
this diagram demonstrates that it is impossible to separate the two
kinematic populations on the basis of their colours and magnitudes.

\section{Discussion}

\subsection{The link with Orion OB1}

The RV distribution shown in Fig.~\ref{rvhist} is clearly bimodal.
Kinematic groups 1 and 2 have weighted mean RVs of $(23.8\pm 0.2)$~\kms\ and
$(31.0\pm 0.1)$~\kms\ respectively (with an external error of $\pm
0.5$~\kms).  The standard deviations about these mean velocities are
1.1~\kms\ and 1.3~\kms, suggesting coherent kinematic groups with
internal dispersions similar to those found in young open
clusters. Previous RV investigations of stars around $\sigma$~Ori had
insufficient precision ($\sigma \simeq 5$~\kms) to properly resolve
these two components (e.g. Walter et al. 1997; Kenyon et
al. 2005). From Figs.~\ref{spatial}, \ref{na} and \ref{cmd} we deduce
that: (i) the two kinematic components have a different spatial
distribution; (ii) on average, group 1 is older than group 2 but by an
uncertain amount and they may overlap; and (iii) the two groups occupy
similar positions in the CMD.

The early-type stars of the Orion OB1 association were originally split
into four subgroups by Blaauw (1964), primarily on the grounds of
position.  The two subgroups relevant to this study are the 1a and 1b
subgroups -- OB1a is an older dispersed population mainly to the north
and north-west of the belt stars, while OB1b consists of the belt region
including the belt stars and $\sigma$ Ori. Brown et al. (1994, 1998)
have reviewed the subgroup properties using new photometry and
Hipparcos parallaxes for the O and B stars. They find the subgroups can
also be separated on the basis of age and distance, obtaining ages of
$11.4\pm 1.9$\,Myr and $1.7\pm 1.1$\,Myr and distances of $326\pm
16$\,pc and $439\pm 33$\,pc for OB1a and OB1b respectively. The two
subgroups cannot be distinguished kinematically. Proper motions are
very small in Orion OB1 because the association is moving almost
radially away from the Sun. Average radial velocities of
$(23.8\pm0.7)$~\kms\ and $(23.1\pm1.4)$~\kms\ are quoted for subgroups
OB1a and OB1b by Morrell \& Levato (1991) based on measurements of large
numbers of early-type stars. Brice\~no et al. (2005) have searched for
low-mass PMS stars in the Orion OB1a and OB1b regions using variability as an
identifying characteristic and spectroscopy for confirmation. Using
Blaauw's spatial criteria to define the subgroups and assuming the
distances quoted above, they compare their data with low-mass
isochrones, finding ages of 7--10\,Myr and 4--6\,Myr for OB1a and OB1b
respectively. The difference in age is supported by the frequency with
which accretion discs are found in the two populations.

All of our targets are spatially within Blaauw's OB1b region,
but given that subgroup OB1a is in the foreground and widely dispersed,
there seems little reason to suppose there are not significant numbers
of older OB1a members in the line of sight towards OB1b (e.g. see
Fig.~9 in Sherry et al. 2004). Hence, based on the similarity of their
mean RVs, kinematic group 1 could be identified with either the OB1a or
OB1b subgroups. Conversely, kinematic group 2 cannot be identified with
any of the Orion OB1 subgroups and has a mean RV which is similar to
the RV of $\sigma$~Ori itself. $\sigma$ Ori AB has $V_{r}=29.9\pm
1.6$\kms\ according to Barbier-Brossat \& Fignon (2000), who analyse
all pre-1990 RV data. Morrell \& Levato (1991) obtain $27\pm 4$\kms\
for $\sigma$~Ori E, but caution that it is probably variable. Given
this kinematic evidence it is hard to see (other than by a simple
positional criterion) how $\sigma$~Ori could be securely classed as
part of Orion OB1b and therefore it should not be assumed that either
$\sigma$~Ori or the group~2 objects around it have a distance and age
in common with Orion OB1b.

Figure~\ref{cmd} shows two isochrones taken from the models of Baraffe
et al. (1998) and converted into the observational plane using the
empirical colour-effective temperature relationship and bolometric
corrections described in Jeffries et al. (2004). These isochrones are
approriate for our current knowledge of the age and distance of the
Orion OB1a and OB1b subgroups. Our targets, along with Orion OB1a
and OB1b, lie close to a locus (or more correctly a band) in the
age-distance plane that results in young coeval populations appearing
in roughly the same position in the colour-magnitude (and
Hertzsprung-Russell) diagram.

The RV evidence, the strong spatial concentration of low-mass PMS stars
and the low ratio of group 1 to group 2 objects immediately around
$\sigma$~Ori strongly support the identification of group 2 as the
``$\sigma$~Ori cluster''.  The Hipparcos parallax of $\sigma$~Ori only
constrains the distance to this cluster to be 270--520\,pc, with a
corresponding age from low-mass isochrones of between 15--2\,Myr.
However, if kinematic group~1 is older on average than group 2 (as
suggested by Fig.~\ref{na}) and, identified with Orion OB1a at $\simeq
10$\,Myr, then this would favour an age of a few Myr for group 2 and a
distance similar to that of Orion OB1b ($\simeq 440$\,pc).  This would
be consistent with a likely main-sequence lifetime of $<7$\,Myr for
$\sigma$~Ori A (O9.5V, $T_{\rm eff}\simeq 3.0\times10^4$\,K -- Schaller
et al. 1992).  Alternatively, if group~1 is
dominated by members of the more youthful Orion OB1b, then $\sigma$~Ori
and the group~2 objects that surround it would have to be even younger
and further away.

There are two ways to further constrain this problem. One would be to
improve the parallax measurement to $\sigma$~Ori and the second would
be to investigate the two kinematic groups with distance-independent
age indicators. Of the latter there are two possibilities.
Photospheric lithium is rapidly depleted in low-mass stars beginning at
an age which is mass (and hence temperature dependent).
Zapatero-Osorio et al. (2002b) found that all their $\sigma$~Ori
association targets across a wide mass range had Li consistent with no
depletion, leading to an upper age limit of 8\,Myr.  Unfortunately
their targets do not have precise RVs and few are contained within our
survey, so we do not know to which kinematic group they belong.
Instead we have cross-correlated members of kinematic groups 1 and 2 with the
targets of Kenyon et al. (2005), finding 39 correlations (10 in group 1
and 29 in group 2). Li-rich stars are found at $R-I\geq 1.6$ in both
groups (warmer stars were not observed by Kenyon et al.), corresponding
to an effective temperature of $\simeq 3400\,K$. Using the models of
Baraffe et al. (1998), this places an upper limit of $<15$\,Myr to the
age of {\it both} groups. A similar limit is found using other evolutionary
models.  Detailed Li spectroscopy of warmer members of groups 1 and 2
might be able to improve these constraints.

A second empirical method is to compare the fraction of objects with
circumstellar material with those in other clusters and assume that
disc dissipation occurs on a universal timescale. We have correlated
objects in kinematic groups 1 and 2 with objects observed at
$JHK_sL'$ by Oliveira et al. (2004, 2006). We find 29 correlations
and that 1/4 members of group 1 and 10/25 members of group 2 have a
significant $K_s-L'$ excess (as defined in those papers). There are
insufficient statistics to comment on the relative ages of the two
groups, but a 40 per cent disc frequency for group 2 is consistent with
an age of 3-4\,Myr according to Haisch, Lada \& Lada (2001).

In summary, our favoured, although not necessarily unique,
interpretation is that kinematic group 1 should be identified with
Orion OB1a at an age of $\simeq 10$\,Myr and at a distance of $\simeq
330$\,pc. The spread about a single isochrone could be due to the
presence of some younger Orion OB1b members and perhaps a large spread
of distances within the widely dispersed OB1a association. Group 2
would then have an age of about 3\,Myr, be clustered around
$\sigma$~Ori and be at a similar distance to Orion OB1b ($\simeq
440$\,pc). However, the ``$\sigma$ Ori cluster'' appears to be
kinematically distinct from Orion OB1b.

\subsection{Consequences}

Studies of low-mass stars and brown dwarfs in the $\sigma$~Ori region
must account for the possibility of a population with mixed age
and distance. These populations cannot be distinguished in the CMD and
the only way to reliably attribute stars to one group or the
other is via RV measurements with a precision better than a few
\kms. The populations do show some spatial segregation. It appears that
samples restricted to within 10 arcminutes of $\sigma$~Ori would
suffer contamination from the older population at the level of only
$\simeq 10$ percent. However, in the quest for larger and more complete
samples, many studies in the literature have travelled beyond this
boundary, into regions (e.g. $>15$ arcminutes to the north of
$\sigma$~Ori) with roughly equal numbers from both
populations. How much of a problem would older interlopers be? There is
insufficient space in this letter to explore in detail the effects of our
discovery on all the previous studies of the $\sigma$~Ori region but we
make some general points below.

The effects on mass determination depend on which technique is used. A
simple mass-magnitude relationship obtained from models (e.g. Baraffe
et al. 1998) is not greatly affected by assignment of a star to one
kinematic group or the other -- the differences in age and distance
almost compensate. Masses for an age of 3\,Myr and distance of 440\,pc
are larger than those deduced for an age of 10\,Myr and distance of
330\,pc by about 6 per cent at $0.5\,M_{\odot}$, by 30 per cent at
$0.06\,M_{\odot}$ and are almost the same at $0.02\,M_{\odot}$.
However, bigger discrepancies would arise if masses were estimated from
colours or spectral types because the evolutionary tracks are not
vertical when using these parameters as abcissae in the
Hertzsprung-Russell diagram.  At a given colour (in the $I$ versus
$I-J$ CMD for example) one might deduce that a star had
$0.5\,M_{\odot}$ by assuming an age of 3\,Myr, but it would actually be
$0.25\,M_{\odot}$ if the age were 10\,Myr (Baraffe et al. 1998). This
discrepancy reduces towards lower masses as the evolutionary tracks
become more vertical in the CMD, becoming a factor of 1.2 at
$0.07M_{\odot}$ and $\simeq 1$ below $0.04\,M_{\odot}$.  There are also
significant structural changes for low-mass stars on these timescales
which would be important if using a $\sigma$~Ori sample to study the
evolution of angular momentum and rotation rates. Under the assumptions
of negligible angular momentum loss between 3 and 10\,Myr and
homologous contraction, objects thought to be at 3\,Myr but which were
actually at an age of 10\,Myr, would have spun up by a factor of 2.2 at
$0.5\,M_{\odot}$ and 2.9 at $0.07\,M_{\odot}$ .  The interlopers could
therefore have a significantly different rotation distribution, perhaps
appearing as a tail of short period, fast rotators.

If the disc lifetime distribution is independent of environment then
interlopers from an older population would have significantly fewer
(accretion) discs and result in a downwardly biased estimate of the
disc frequency in the $\sigma$~Ori cluster. As discussed in the
previous section there are insufficient data to test this in detail
yet. The disc frequency we determined for members of kinematic group 2
is higher than, but still consistent with, the $33\pm 6$ per cent
frequency determined for a larger sample of objects around $\sigma$~Ori
which contains at least some objects from group 1. 

Finally, studies of
the age spread and spatial distribution within the cluster must
be aware of, and try to eliminate, the effects caused by including two
populations with probable differences in age and distance. For example,
the radial profile of the cluster deduced by Sherry et al. (2004)
includes photometrically selected cluster members more than 30
arcminutes to the north of $\sigma$~Ori, which are quite likely to
belong to kinematic group 1. Excluding these would reduce the number
of stars and brown dwarfs in the $\sigma$~Ori cluster and decrease the
deduced cluster core radius.

\section{Conclusions}

The cluster of young objects situated towards $\sigma$~Ori consists of
two kinematic components separated by $\simeq 7$\,\kms\ in RV. The
first component (and larger in our sample) is concentrated around
$\sigma$~Ori and shares a common RV with it. The second component
increases in concentration towards the north and has a RV similar to
the Orion OB1a and OB1b sub-associations. On the basis of
gravity-sensitive spectral features, the second component appears to be
older on average than the first, although considerable overlap is
possible. The two components cannot be separated in a colour-magnitude
diagram, a degeneracy that implies that the older second component is
closer. Our favoured interpretation is that the first component is the
``$\sigma$~Ori cluster'', but that it should not necessarily be
identified (in distance or age) with any part of the Orion OB1
association. The second component consists of low-mass objects from
either the Orion OB1a or OB1b sub-associations or more probably, a
mixture of the two. The presence of multiple populations with different
ages complicates the study of low-mass objects in this region.

\section{Acknowledgements}
Based on observations collected with the VLT/UT2 Kueyen telescope
  (Paranal Observatory, ESO, Chile) using the FLAMES/GIRAFFE
  spectrograph (Observing run 076.C-0145).

\nocite{rothman05}
\nocite{schaller92}
\nocite{walter97}
\nocite{haisch01}
\nocite{jeffries04}
\nocite{bejar01}
\nocite{zapatero02}
\nocite{zapatero02b}
\nocite{zapatero00}
\nocite{kenyon05}
\nocite{sherry04}
\nocite{brownorion94}
\nocite{brown98}
\nocite{walter97}
\nocite{oliveira04}
\nocite{oliveira06}
\nocite{baraffe98}
\nocite{perryman97}
\nocite{barradosigori03}
\nocite{franciosini06}
\nocite{caballero04}
\nocite{blaauw64}
\nocite{burningham05}
\nocite{montes98}
\nocite{xu91}
\nocite{barbier-brossat00}
\nocite{morrell91}
\nocite{briceno05}
\nocite{schiavon97}
\nocite{bailer-jones04}
\nocite{burningham05b}
\nocite{jayawardhana03}
\nocite{nicholls88}

\label{lastpage}
\bibliographystyle{mn2e}  
\bibliography{iau_journals,master}

\end{document}